# A new method for direct rf power absorption studies in CMR materials and high $T_c$ superconductors

S. Sarangi and S. V. Bhat[*]

Department of Physics, Indian Institute of Science, Bangalore – 560012, India

[*]Corresponding author:

    S. V. Bhat

    Department of Physics

    Indian Institute of Science

    Bangalore – 560012, India

    Tel.: +91-80-22932727, Fax: +91-80-23602602

    E-mail: svbhat@physics.iisc.ernet.in

**Abstract:**

The design, fabrication and performance of an apparatus for the measurement of direct rf power absorption in colossal magnetoresistive (CMR) and superconducting samples are described. The system consists of a self-resonant *LC* tank circuit of an oscillator driven by a NOT logic gate. The samples under investigation are placed in the core of the coil forming the inductance *L* and the absorbed power is determined from the measured change in the current supplied to the oscillator circuit. A customized low temperature insert is used to integrate the experiment with a commercial Oxford Instruments cryostat and temperature controller. The oscillator working in the rf range between 1 MHz to 25 MHz is built around an IC 74LS04. The temperature can be varied from 4.2 to 400 K and the magnetic field from 0 to 1.4 T. The apparatus is capable of measuring direct power absorption in CMR and superconducting samples of volume as small as $1 \times 10^{-3}$ cm$^3$ with a signal to noise ratio of 10:1. Further increase in the sensitivity can be obtained by summing the results of repeated measurements obtained at a given temperature. The system performance is evaluated by measuring the absorbed power in La$_{0.7}$Sr$_{0.3}$MnO$_3$ (LSMO) CMR manganite samples and superconducting YBa$_2$Cu$_3$O$_7$ (YBCO) samples at different rf frequencies. All operations during the measurements are automated using a computer with a menu-driven software system, user input being required only for the initiation of the measurement sequence.

**Introduction:**

Recent years have seen the development of two new exciting classes of materials in condensed matter science, mainly colossal magnetoresistive (CMR) manganites and the high $T_c$ superconductors. Both of these type of materials are characterized by strongly correlated electronic systems in them, making the understanding of the phenomena a challenging task. At the same time they also promise enormous practical applications and therefore the measurement of their physico-chemical properties is also very important. From the device applications point of view, one of the crucial parameter is the power dissipation, especially when subjected to AC fields, which needs to be minimized. It turns out that both in CMR manganites and high $T_c$ superconductors, the ac losses, especially in the rf and microwave ranges can have both magnetic and transport contributions. The high $T_c$ materials are type II superconductors with low $H_{c1}$ (lower critical field) values and thus allow penetration of even low magnetic fields in the form of quantized flux lines. The motion of these flux lines in response to the induced rf current leads to power dissipation. In addition, they being intrinsically granular, there is an additional power loss mediated by the Josephson junctions. In the case of CMR manganites the spin, charge and structural degrees of freedom are intimately coupled and magnetic and transport properties are inseparable from each other. While from a theoretical point of view it would be informative to be able to determine the relative contributions of magnetic and transport losses separately, for most applications it suffices to determine the absolute and total magnitude of power dissipation.

Different techniques have been used in the past to study the dissipation behavior of high $T_c$ materials and CMR manganites. One such method especially used for the study of superconductors are the technique of non-resonant rf power absorption (NRRA) and non resonant microwave absorption (NRMA). Here conventional CW NMR (Nuclear Magnetic Resonance) and EPR (Electron Paramagnetic Resonance) spectrometers are used to record the non resonant response of the sample as a function of temperature and magnetic field. Magnetic field modulation and phase sensitive detection make this a highly sensitive electrodeless technique for the characterization of even miniscule a moment of the samples. However, the main drawback of this technique is its inability to provide the absolute power loss since only the magnetic field derivative of the absorbed power (dP/dH) is recorded.

In this work we present a method to measure the absolute power absorbed when either the CMR or the superconducting sample is subjected to the ac field. We make use of a NOT logic gate based oscillator, the change in the current supply to which gives a measure of the absorbed power. Such oscillators have earlier been used [1, 2], but focusing on the changes in the frequency of the oscillator, which was used to extract information on the change in the penetration depth and the occurrence of various transitions. As discussed in the following sections we extend the capability of the technique to include the measurement of ac dissipation, an important parameter for the characterization of materials for device applications.

**Operating Principle and Circuit Design:**

The technique involves placing the sample in the coil which is a part of a *LC* circuit of resonant frequency *f* in the rf range and measuring the change in the total current flow in the circuit. When physical properties of the sample change the rf energy absorbed by the sample also changes which can be a function of temperature, magnetic field, sample orientation or the resonant frequency itself. According to the rf energy absorbed by the sample the total current supply to the circuit changes. At a particular stage the product between the change in current $\Delta I = (I_1-I_0)$ and the supply voltage *V* is the net power absorbed by the sample. Here $I_1$ is the current supply to the oscillator circuit when sample is present inside the coil *L* and $I_0$ is the current when the coil is empty. In our case we keep the voltage *V* always 5 volts for general experiments. We can vary this voltage from 3 to 20 V depending on the rf power needed for the experiment.

The sensitivity and stability of this technique depend completely on the oscillator. The operating principle of a rf oscillator is very simple. The oscillators work on a form of instability caused by a regenerative feedback without which the input dies out due to energy losses. The reactive element in a positive feed back circuit causes the gain and phase shift to change with the frequency. In general, there will be only one frequency corresponding to which the gain is unity and the phase shift is equivalent to $0°/360°$. This satisfies the basic criteria for production of sustained oscillations. An *LC* tank circuit is maintained at a constant amplitude resonance by supplying the circuit with external power to compensate dissipation.

The change in the current is measured with a digital multimeter (Keithley model 2002), which has DC current measurement sensitivity of 10 pAmp. The fluctuations introduced due to electrical and thermal noise is taken care of by measuring the mean of 50 data points. The above measurement can also be done by any sensitive voltmeter. Here we need to convert the current to voltage by passing the current through a known resistance. The resistance should be of a low value so that it does not restrict the supply current to the oscillator. Typically 1 Ω is a good choice for the resistance.

Proper grounding is one of the crucial factors influencing the measurement accuracy. The best way is to create only one ground point to which all instruments are connected. The fewer electronics instruments involved in the setup the better, as different instruments have their own grounds at different potentials. Connection of instruments creates ground loops and ground loop currents. We recommend using the analogue ground of all the instruments and not the line ground. As this setup is very sensitive towards ground problem, it is very important to connect all the metallic part of the cryostat and the helium dewar also to the same ground.

Fig. 1 shows the oscillator circuit using 74LS04, which is a, TTL bipolar Hex inverter. It is easily available, cost effective and devoid of any biasing problems. The current in the inductor and hence the field $H_f$ can be varied by changing the input voltage at pin 14 of the device. 74LS04 contains six NOT gates with typical propagation delay time ≈ 15 ns. This specifies the upper limit for its application as a high frequency oscillator. If the voltage at point "a" is low, then output at 2 (point "b") is high and the current builds up in the inductor, which in turn transfers its energy to the capacitor. Due to the charging of capacitor, voltage at "a" develops to high state and output at "b" becomes low and this leads to sinusoidal oscillations. The frequency of oscillation is determined by series $LC$ circuit and is given by the standard expression $\omega=1/(2\pi\sqrt{(LC)})$.

The tank inductance $L$ and capacitance $C$ are vital components of the Integrated circuit oscillator (ICO) and have to be chosen with utmost care. For $C$, surface mount high-Q rf chip capacitor (American Technical Ceramics) with values ranging from 50 to 2000 pF were tried out. Note that it is important to have the value of $C$ higher than the capacitance of the 1.5-ft RG402 coaxial assembly, which is rated at 29 pF/ft. The inductive coil is a 15-turn solenoid hand-wound using AWG30 insulated Cu wire around a hollow ceramic tube (0.8-cm diameter) and potted with polystyrene epoxy resin. A coating of GE varnish is also applied for good thermal conductivity. Coils with various diameters and wire thickness were tested for stable performance of the tank circuit. Best results were obtained for typical inductance value of $L$=2-9 µH. The output of this circuit is considerable high, typically .7 V with 5 V input and therefore the circuit can be used for gathering information on different type of losses.

**Low Temperature Cryostat:**

The versatility of the experiment is greatly enhanced when it is adapted to conduct measurements on materials over a wide range in temperature, magnetic field and frequency. This is achieved in our system through the integration of our home built circuitry with a customized cryogenic user probe that fits into the commercial Oxford liquid helium cryostat. A semirigid coaxial cable assembly is attached to the probe. The inductive coil $L$ from the circuit shown in Fig. 1, is connected to the bottom end of the coaxial cable while the top end is terminated by a female connector that can mated to the rest of the oscillator circuit. A schematic of the coaxial probe used for the variation in temperature and magnetic field is shown in Fig. 2.

Samples are placed in gelcaps and inserted into the resonant coil. They are securely fastened with Teflon tape to ensure rigidity. The sample and coil are in thermal contact with the temperature sensor (Cernox). Thermal contact is made using Apiezon "N" grease. The coax is fixed in the central bore of the user probe. A double "O" ring seal at the top flange (not shown) is used to maintain a vacuum seal around the coax cable. The inner conductor and the outer shielding are used for electrical connections. Such an arrangement requires only one cable but the coax must be electrically isolated from the surrounding environment. Electrical isolation was obtained by wrapping the coax in Teflon tape and placing a small rubber "O" ring close to the base of the coax. Once the sample and coax were mounted in the user probe, the probe was then inserted into the cryostat. The Oxford temperature controller and the Bruker electromagnet are then used as a platform for varying the sample temperature (*4K<T<400K*) and static magnetic field (*0T<H<1.4T*).

Thus, our design combines the ease of operation of the temperature controller (including the possibility of changing samples without warming up the system) and the static magnetic field control with the oscillator setup.

**Computer Interface and Data Acquisition:**

Control of the temperature, magnetic field and the frequency was through using the standard software and a data acquisition computer. Sequences can be written that would control the temperature and field as a function of time. The analog output ports of a Gauss meter and Cernox censor were utilized as monitors for the field and temperature. The signals were then connected to the 2002-multimeter. Here the 2002- multimeter not only measures the supply current but also measures the DC voltage coming from the Gauss meter for magnetic field measurement and the resistance of the Cernox censor for temperature measurement. We use a scanner card for all the measurements in the 2002-multimeter. It should be noted that this was only used to convert the analog voltage to general-purpose interface bus (GPIB) readable data. This data is read via GPIB interface using the same data acquisition computer. The same computer also communicates with the frequency counter, multimeter, oscilloscope, temperature controller and magnet power supply unit.

A typical run would scan magnetic field, temperature or frequency over a certain range. The data acquisition computer would then monitor the GPIB data bus reading temperature, magnetic field, frequency and the change in current, which is a measure of the power being absorbed. A schematic of this arrangement is shown in Fig. 3.

**System Performance:**

Figure 4 shows the output signal trace from the storage oscilloscope with the IC oscillator at a resonant frequency around 2 MHz. A clean nearly sinusoidal waveform is apparent with peak-to-peak amplitude of around 700 mV. Increasing or decreasing the supply voltage can adjust the amplitude. If we change the supply voltage from 5 to 10 V the rf oscillation amplitude changes from 0.7 to 1.4 V. Changing the inductance $L$ of the coil or the capacitance $C$ of the capacitor changes the resonant frequency.

As mentioned earlier, the experiment depends crucially on the stability of the oscillators. In Fig. 5 we show the stability of the ICO circuit during a typical magnetic field scan. It is taken at 200 K by ramping the field in the steps of 10 Gauss up to 1.2 Tesla and back to zero in 1 hour. The maximum current (45 Amp) corresponds to a dc field of 14 kG for 8.4 cm pole separation. The Fig. 5 also reflects the negligible drift of the oscillations with time. The 4 micro Amp fluctuation is insignificant compared to the magnitude of the effects, that is a current shift of 2.5 mAmp at the superconducting transition (Fig. 9) and 120 micro Amp at the CMR transition (Fig 13) in LSMO. Long-term stability tests of the ICO were conducted and our circuit showed excellent stability. The change due to the drift in the resonant frequency (~ 1 KHz) is limited to 1 micro amp over a period of 20-30 min. Tests over a 24 h period established an overall drift around 5 micro amp (Fig. 6).

The temperature dependence shown in Fig. 7 is a combined result of decrease in resistivity of the Cu wire making up the coil and the effect of thermal contraction, as the temperature is lowered. In Figure 8 we show the power drift with frequency. The frequency is varied with only changing $C$ and keeping the amplitude of rf oscillation constant. This shows very minute change in current with frequency. All the above experiments for system performance were done in the absence of any samples.

It is difficult to track down the precise source for all these drifts as several factors can contribute. Some possible candidates include drift in the bias supply, IC operation, and thermal dissipation inside the enclosing Al box. Nevertheless, these small drifts are at least a few order of magnitudes less than typical current shift encountered in a measurement and do not affect the result. It must be mentioned, however, that an order of magnitude improvement in stability can be achieved by giving proper ground to the metallic part of liquid helium cryostat and the liquid helium dewar.

Both the temperature and field dependence of current shift for the coil are quite repeatable and not very different for the two cases viz. Coil axis parallel or perpendicular to the dc field provided by the magnet. As a precaution, it should be noted that if the coil is not rigidly mounted, one may have to contend with the movement of the coil due to Lorentz force acting on it when the oscillating rf field (and hence the rf current) is perpendicular to the applied static field.

**Measurements and data collection:**

In our first experiment we investigated the high $T_c$ cuprate superconductor YBCO. We prepared two pellets made of polycrystalline samples of YBCO. Both the pellets were exactly similar in size but prepared with different techniques. Both the samples show sharp superconducting transition temperature at ~ 91 K as determined by $\rho \sim T$ and AC susceptibility measurements. The material was found to be single phasic as determined by x-ray diffraction. The samples were placed in the coil. The filling factor of these samples was ~ 0.7. At zero fields the noise in measuring supply current was less than 1 micro Amp. The superconducting transitions are clearly visible in both the sample (Fig. 9 & Fig. 11) with the background (discussed above) subtracted. One sample (Sample 1) shows the rf power absorption is less in the superconducting state than the normal state (Fig. 9) whereas the other sample (Sample 2) shows the power absorption is more in the superconducting state than the normal state (Fig. 11). Fig. 9 shows that in the normal state the supply current value is 9.328 mA and superconducting state the supply current is 6.915 mA. From these values we can easily say that in the normal state, the sample #1 absorbs ~12 mW (the product of supply voltage with the change in current) more energy than its superconducting state. Fig 10 & 12 are the magnetic field dependent power absorption at various temperatures in both the samples. Here we have plotted the relative change in current instead of exact value of current just to visualize the change due to field at different temperature. One sample shows that the rf power absorption increases with increasing field whereas the other one shows the power absorption decreases with increasing field. Here it is necessary to note that an increase in current value means that the rf oscillations are damping very fast and to sustain the oscillations more current has to be supplied. So more power absorption leads to more supply current to the oscillator. For sample #1 the power absorption smoothly and monotonically increases with increasing applied field and shows a tendency to saturate at higher fields. We ascribe this change to be directly associated with the Joshepson Junction decoupling [13] and fluxon motion [14] in high $T_c$ superconductors. The shape of the curves and the opposite behavior in these two samples are associated with flux creep, flux flow, Joshepson Junction critical current, number density of Joshepson Junctions and the applied frequency. Detailed interpretation of the result and a comparison of preparation techniques which leads to different type of power absorption are beyond the scope of this instrumentation article and will be discussed in a forthcoming publication [15].

In our second experiment we investigated with manganite sample. The ICO based NRRA measurements on the manganite sample, $La_{0.7}Sr_{0.3}MnO_3$ (LSMO), are presented in Fig. 13 and 14. This system was recently synthesized and it's magnetic and magnetotransport characteristics were studied at our institute. The sample undergoes a paramagnetic to ferromagnetic transition as the temperature is lowered with the Curie temperature $T_c$ around 362 K. In the ICO experiment, the LSMO material is pelletized to capsule form so that it fits snugly into the core of inductive coil. Figure 13 shows the temperature dependence of the power absorbed in zero field. Note that the change in the current is much larger than the background shifts due to empty coil and this has been subtracted in the data presented in Figs. 13 and 14. The paramagnetic to ferromagnetic transition is distinctly seen (Fig. 13) and the data are consistent with the existing reports on this material [16, 17]. In the paramagnetic state the sample absorbs more energy than its ferromagnetic state. The field dependence for the same sample is plotted in the Fig. 14 where we have shown the data for two cases where the temperature is held above and below $T_c$. The striking difference in the current variation in the paramagnetic and the ferromagnetic phase of the sample is quite obvious. For $T>T_c$, the current value smoothly and monotonically decreases with increasing applied field and shows a tendency to saturate at higher fields. We ascribe this change to be directly associated with the magneto impedance (MI), which is dominated by rapid change in the permeability. The shape of the curve for $T<T_c$ and the saturation field are different from that seen in the curve for $T>T_c$. For $T<T_c$, the magnetic field dependent power absorption goes through a peak and then saturates. Detailed interpretation of the results and a comparison of the power absorption for these systems are beyond the scope of this instrumentation article and will be discussed in a forthcoming publication [18].

**Discussion:**

We have demonstrated a contactless method for measuring magnetic and trasport properties of materials at different temperatures and magnetic fields by using this ICO based techniques. Many improvements will enhance this ICO method in future. Use of tunnel diode oscillator in place of IC oscillator would give better stability and resolution. The technique can be useful for the calculation of ac self-field loss in the rf region. Preliminary results on CMR and superconducting samples indicate that this instrument provides a novel way to study the spin and charge dynamics in CMR materials and Josephson Junctions and vortex phenomena in superconducting materials.

**Acknowledgments:**
SVB would like to thank CSIR, UGC and DST, India for financial support.

**Figure Legends:**

1. RF oscillator circuit with IC 74LS04.

2. Schematic of the low temperature probe showing the sample region.

3. Layout of the complete ICO measurement system displaying computer control and data acquisition stages.

4. The output waveform from the ICO circuit measured with a digital oscilloscope. The resonant frequency is around 2 MHz.

5. The Current passing in the circuit at 200 K plotted against magnetic field for the empty coil. The duration of field scan is 1 h.

6. The current passing in the circuit against time when the coil is empty. This is a full day experiment.

7. The current as a function of temperature for the empty coil.

8. The current passing in the circuit at room temperature against resonant frequency of the oscillator.

9. Measured zero-field temperature dependence of the current for the YBCO (sample 1). The critical temperature $T_c$ is 91 K.

10. Field dependence of current for the YBCO (sample 1) above and below the critical temperature $T_c$. The magnetic field is scanned from −150 Gauss to +150 Gauss in 60 seconds.

11. Measured zero-field temperature dependence of the current for the YBCO (sample 2). The critical temperature $T_c$ is 91 K.

12. Field dependence of current for the YBCO (sample 2) above and below the critical temperature $T_c$. The magnetic field is scanned from −150 Gauss to +150 Gauss in 60 seconds.

13. Measured zero-field temperature dependence of the current for the LSMO sample. The paramagnetic to ferromagnetic transition at 362 K.

14. Field dependence of current for the LSMO sample above and below the ferromagnetic transition temperature $T_c$. The magnetic field is scanned from −150 Gauss to +150 Gauss in 60 seconds.

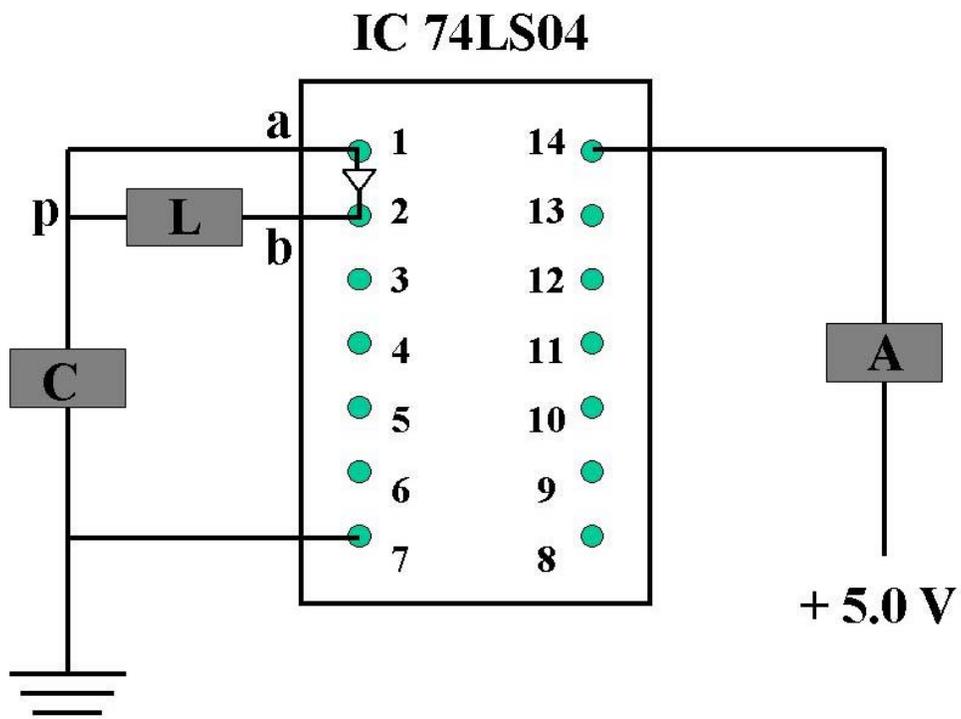

**FIG. 1**

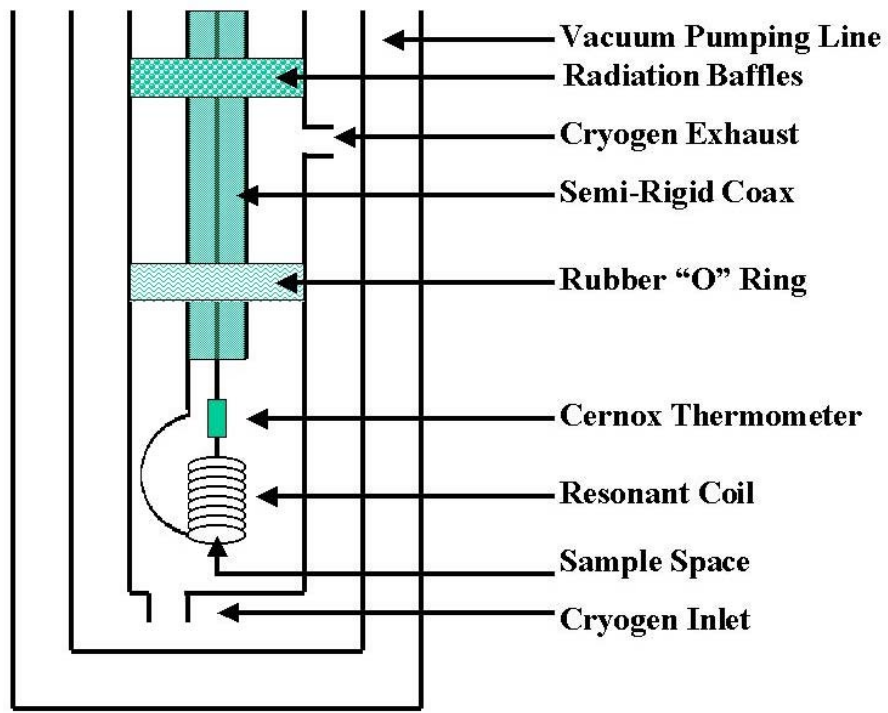

**FIG. 2**

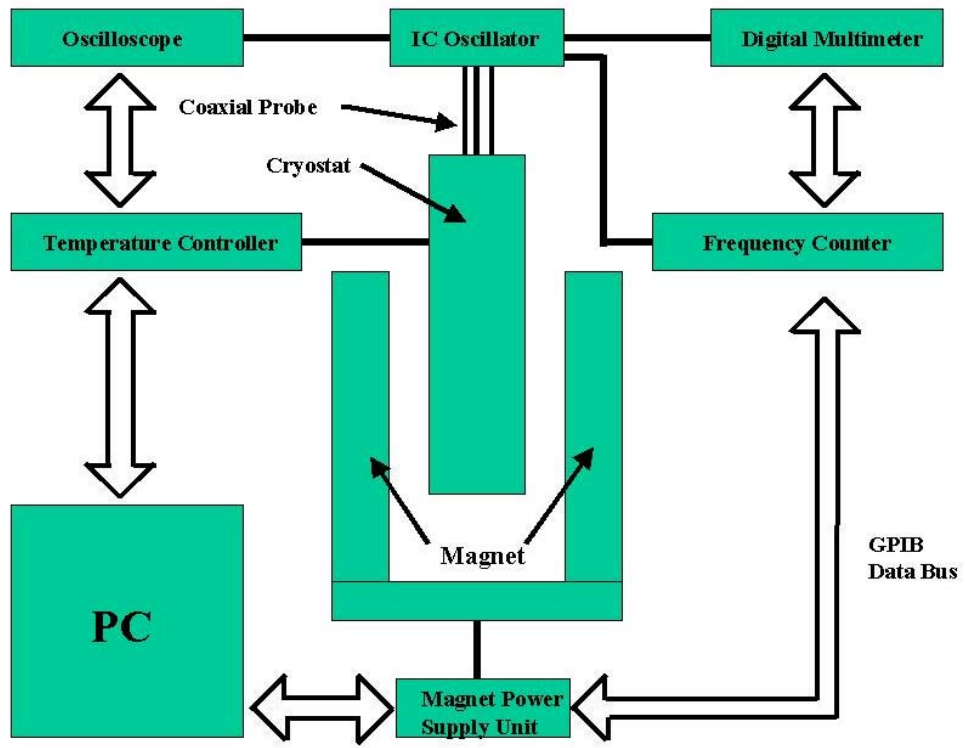

**FIG. 3**

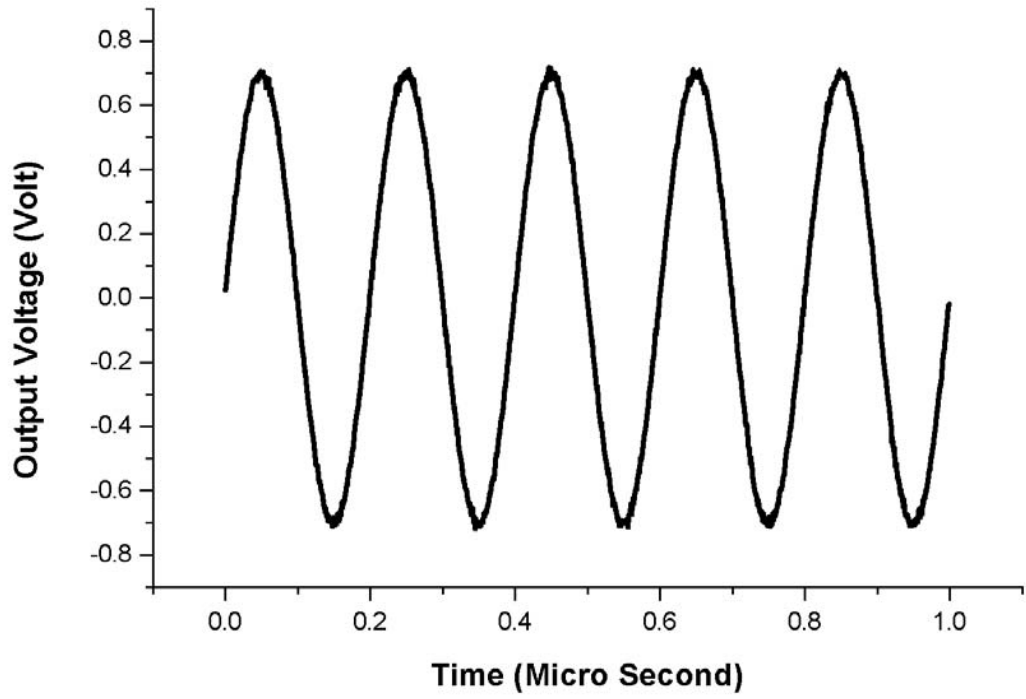

**FIG. 4**

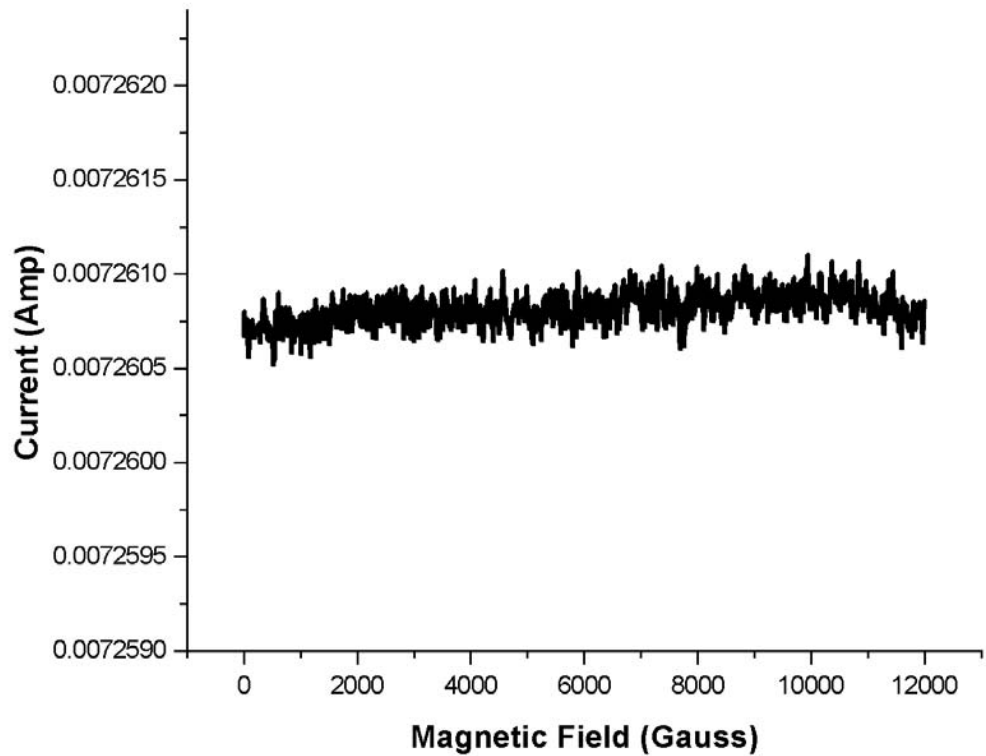

**FIG. 5**

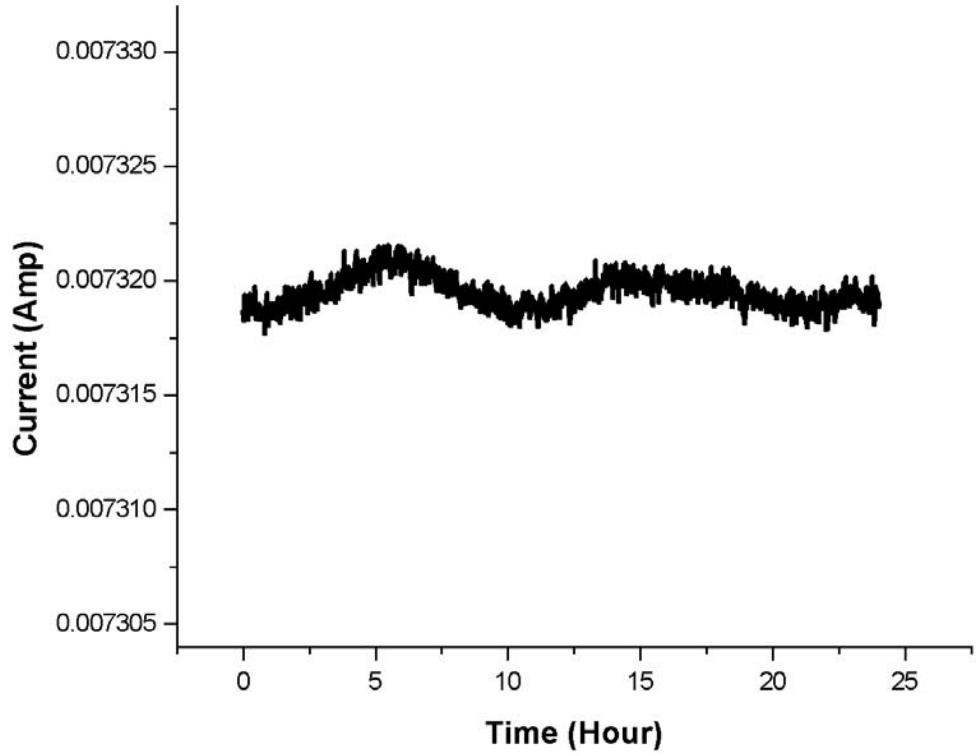

**FIG. 6**

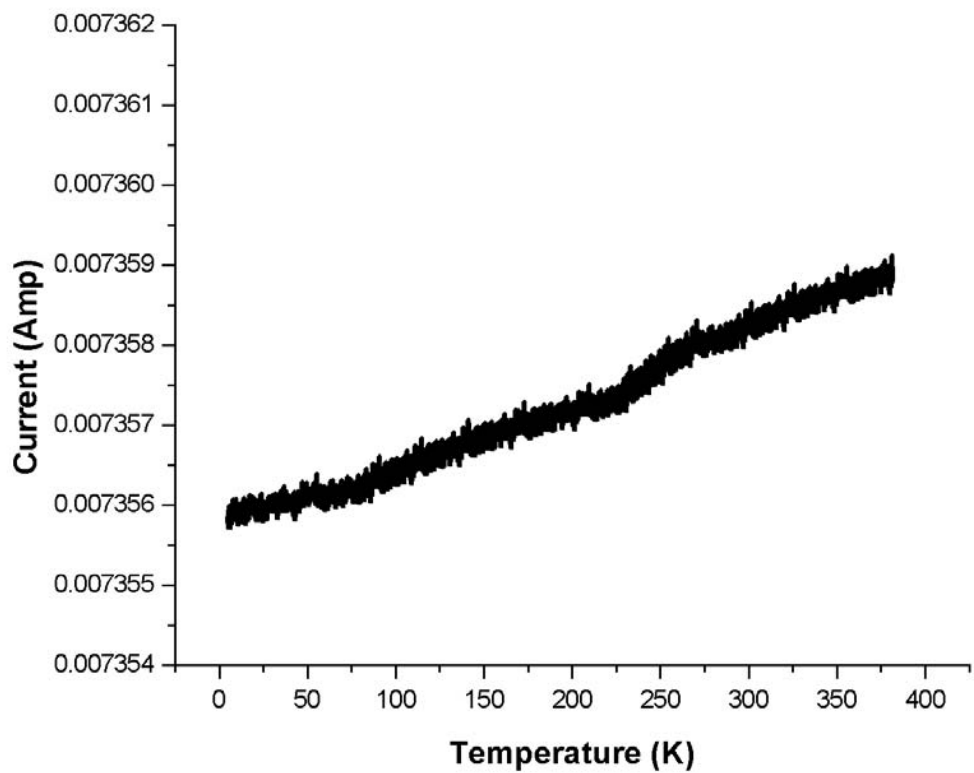

**FIG. 7**

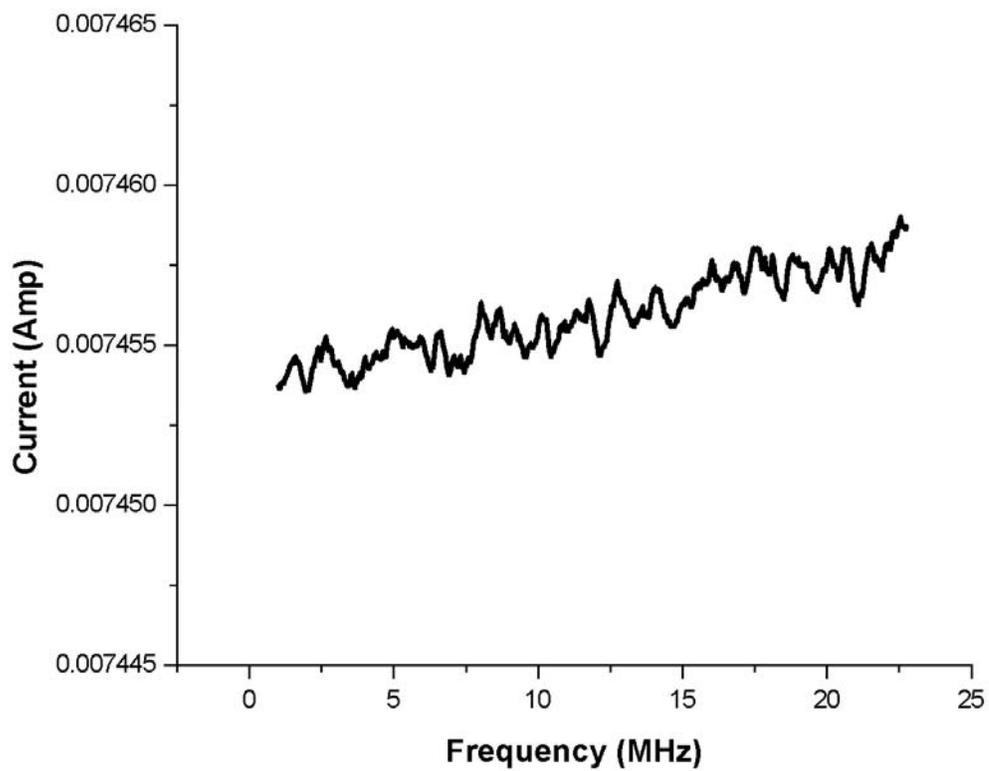

**FIG. 8**

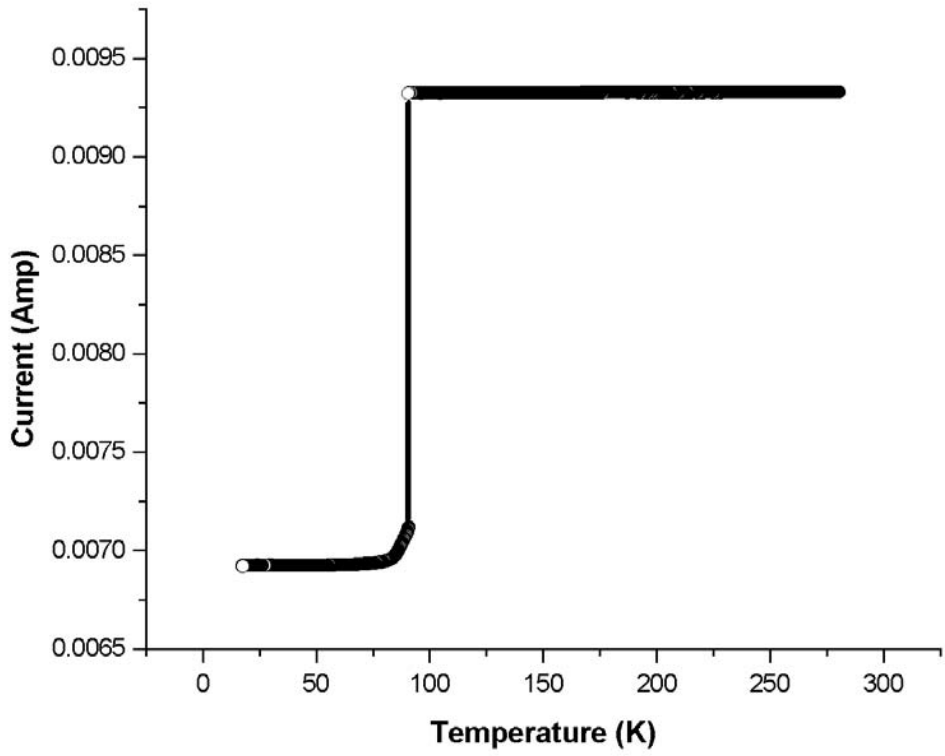

**FIG. 9**

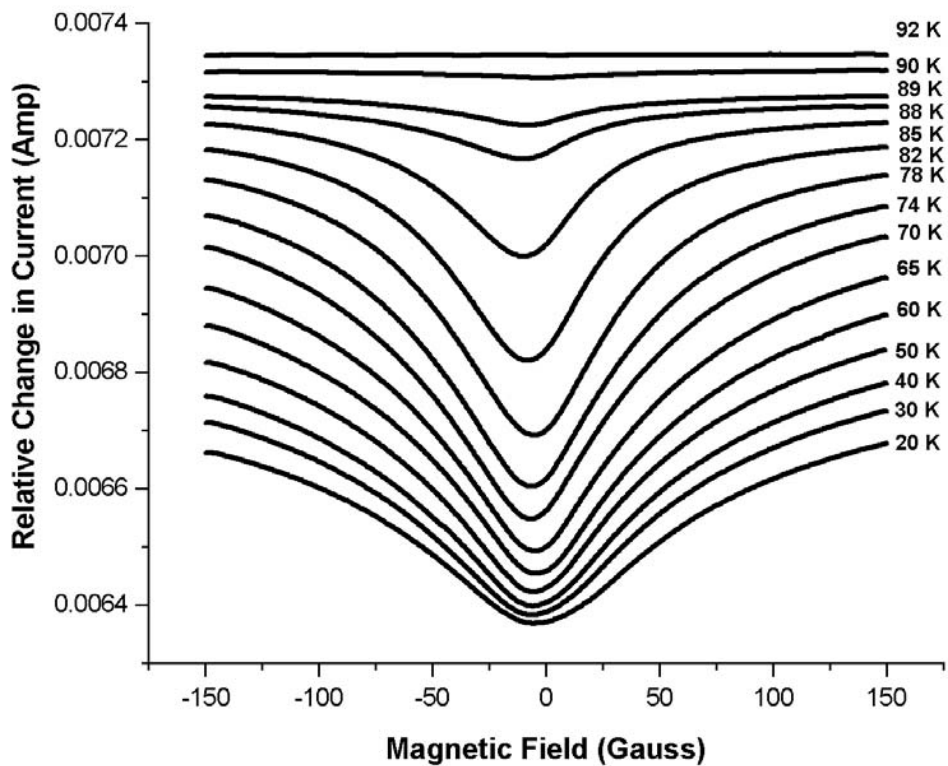

**FIG. 10**

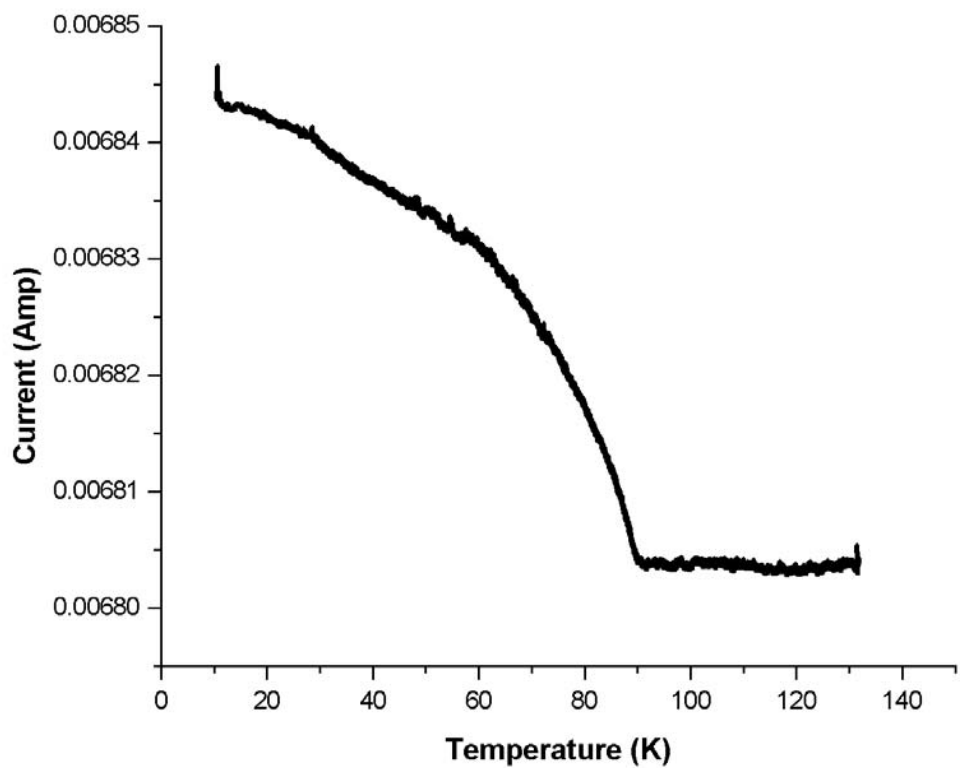

**FIG. 11**

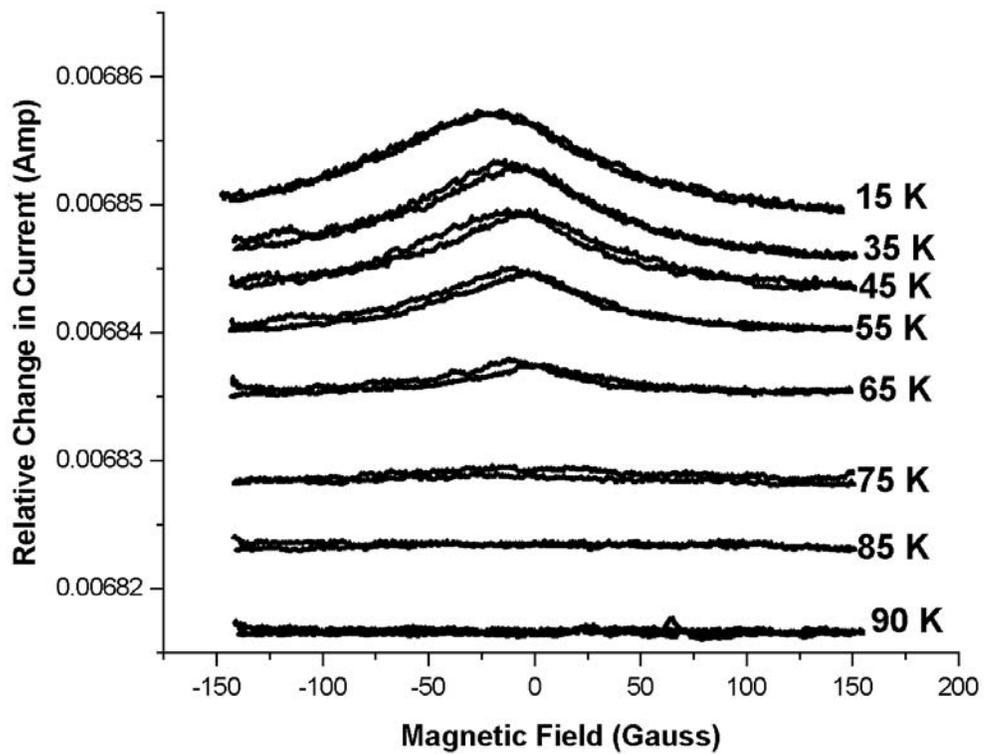

**FIG. 12**

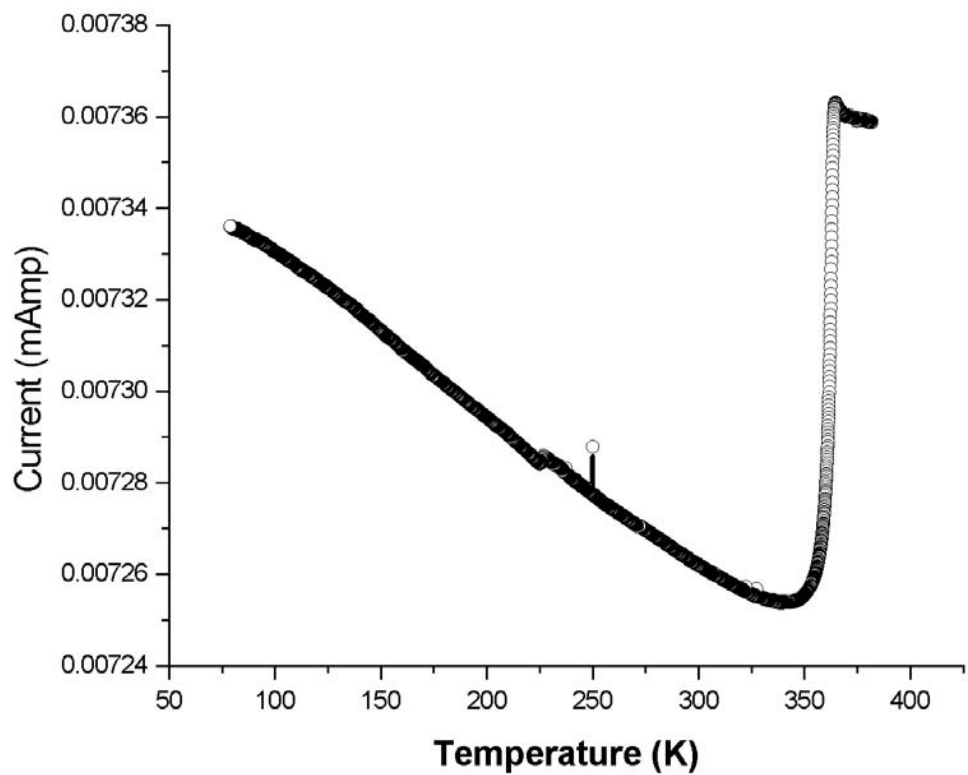

**FIG. 13**

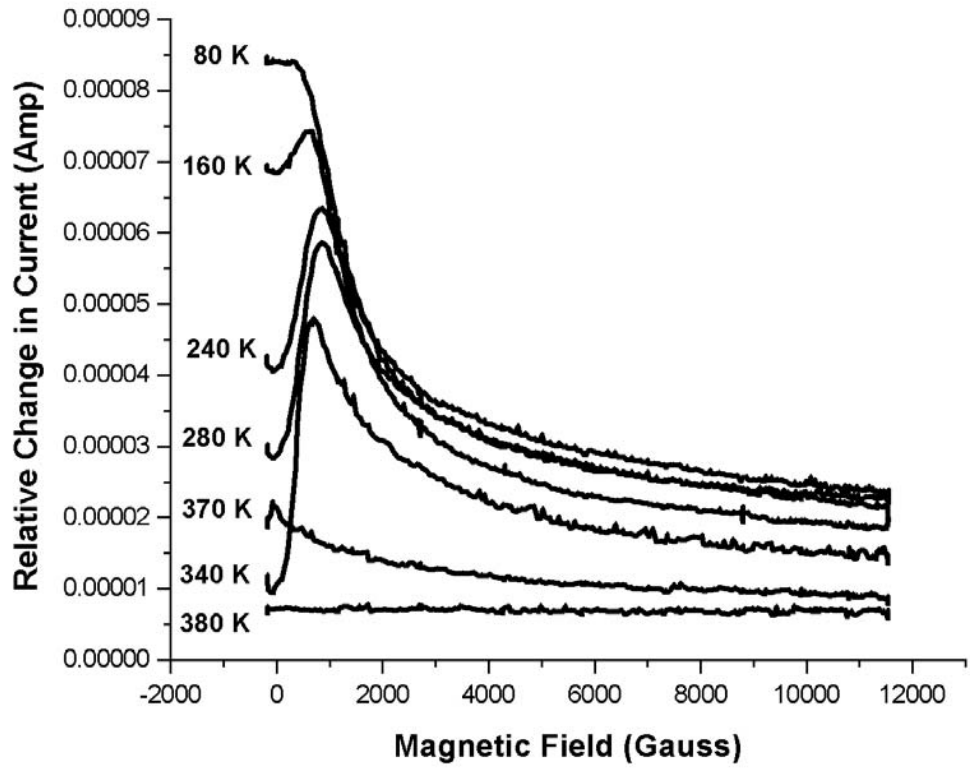

**FIG. 14**